\documentclass[5p]{elsarticle}

\usepackage{lineno,hyperref}
\usepackage{gensymb}
\modulolinenumbers[5]

\journal{Journal of \LaTeX\ Templates}

\begin{document}

%\bibliography{CHIPTRAPrefs}

\begin{frontmatter}

\title{Status and Outlook of CHIP-TRAP: the Central Michigan University High Precision Penning Trap}

\author[CMU,NSCL]{M.~Redshaw\corref{cor1}}
\ead{redsh1m@cmich.edu}
\author[CMU]{R.A.~Bryce}
\author[CMU]{P.~Hawks}
\author[CMU]{N.D.~Gamage}
\author[CMU]{C.~Hunt}
\author[CMU]{R.M.E.B.~Kandegedara}
\author[CMU]{I.S.~Ratnayake}
\author[CMU]{L.~Sharp}

\cortext[cor1]{Corresponding author}

\address[CMU]{Department of Physics, Central Michigan University, Mount Pleasant, Michigan 48859, USA}
\address[NSCL]{National Superconducting Cyclotron Laboratory, East Lansing, Michigan 48824, USA}
%%\address[MSU]{Department of Physics and Astronomy, Michigan State University, East Lansing, Michigan 48824, USA}
%%\address[FRIB]{Facility for Rare Isotope Beams, East Lansing, Michigan 48824, USA}

\begin{abstract}
At Central Michigan University we are developing a high-precision Penning trap mass spectrometer (CHIP-TRAP) that will focus on measurements with long-lived radioactive isotopes. CHIP-TRAP will consist of a pair of hyperbolic precision-measurement Penning traps, and a cylindrical capture/filter trap in a 12 T magnetic field. Ions will be produced by external ion sources, including a laser ablation source, and transported to the capture trap at low energies enabling ions of a given ${m/q}$ ratio to be selected via their time-of-flight. In the capture trap, contaminant ions will be removed with a mass-selective rf dipole excitation and the ion of interest will be transported to the measurement traps. A phase-sensitive image charge detection technique will be used for simultaneous cyclotron frequency measurements on single ions in the two precision traps, resulting in a reduction in statistical uncertainty due to magnetic field fluctuations.
\end{abstract}


\end{frontmatter}

\section{Introduction}
Over the last few decades the use of Penning traps for precise atomic mass determinations of both stable and radioactive isotopes has provided a wealth of data for a wide range of applications in atomic, nuclear, and neutrino physics, astrophysics, and for tests of fundamental symmetries, see e.g. Ref. \cite{Blaum2006} for a recent review. 

Two main measurement techniques for high-precision Penning trap mass spectrometry (PTMS) exist: time-of-flight ion cyclotron resonance (TOF-ICR) and image charge (IC) detection. TOF-ICR is a destructive technique well suited to measurements with short-lived isotopes at on-line PTMS facilities where externally produced ions are transported to the Penning trap at low energies and are dynamically captured. Fractional precisions of $\sim$10$^{-7}$ to 10$^{-9}$ are typically obtained, which are adequate for applications related to nuclear structure, astrophysics, and tests of fundamental symmetries. The recently developed related technique, phase imaging ion cyclotron resonance (PI-ICR), has achieved precisions of $\sim$10$^{-10}$ \cite{Eliseev2013}. IC detection is a non-destructive technique for which precisions of $\sim$10$^{-10}$ to 10$^{-11}$ can be achieved, but for which long measurement times are required . Penning traps utilizing the IC detection technique typically produce ions inside the trap by ionizing background gas or vapor allowed to enter the trap \cite{Shi2005,Dyck2004}. This could limit the ability to perform measurements with long-lived radioactive isotopes that are only available in small quantities. 

There are a number of physics cases of current interest for which ultra-high precision mass determinations of long-lived isotopes are required. For example, a measurement of the $^{163}$Ho (t$_{1/2}$ = 4570 yr) electron capture (EC) Q-value is important for experiments that aim to perform a direct measurement of the electron neutrino mass using calorimeters to study the de-excitation energy spectrum of the daughter, $^{163}$Dy\cite{Gastaldo2014,Alpert2015,Croce2014}. The $^{163}$Ho EC Q-value has recently been measured using PI-ICR to a precision of $\sim$30 eV \cite{Eliseev2015}. However, a precision of $\sim$1 eV (the expected resolution of the calorimeters), corresponding to $\sim$7 $\times 10^{-12}$ in the cyclotron frequency ratio, is ultimately required.
The $^{36}$Cl (t$_{1/2}$ = 3.01 $\times$ 10$^{5}$ yr) neutron separation energy, which can be obtained from a measurement of the $^{36}$Cl $-–$ $^{35}$Cl mass difference, is of interest for a direct test of $E = mc^{2}$. The mass measurement can be compared with high-precision spectroscopy measurements of $\gamma$-rays emitted by $^{36}$Cl after cold neutron capture on $^{35}$Cl \cite{Dewey2006}. The $\gamma$-ray spectroscopy measurements have been performed to a precision of 1.8 eV \cite{Krempel2010}, which corresponds to $\sim$5 $\times 10^{-11}$ in the cyclotron frequency ratio.

\section{Experimental set-up}

\subsection{Overview}
\begin{figure}
\centering
\includegraphics[scale=0.6]{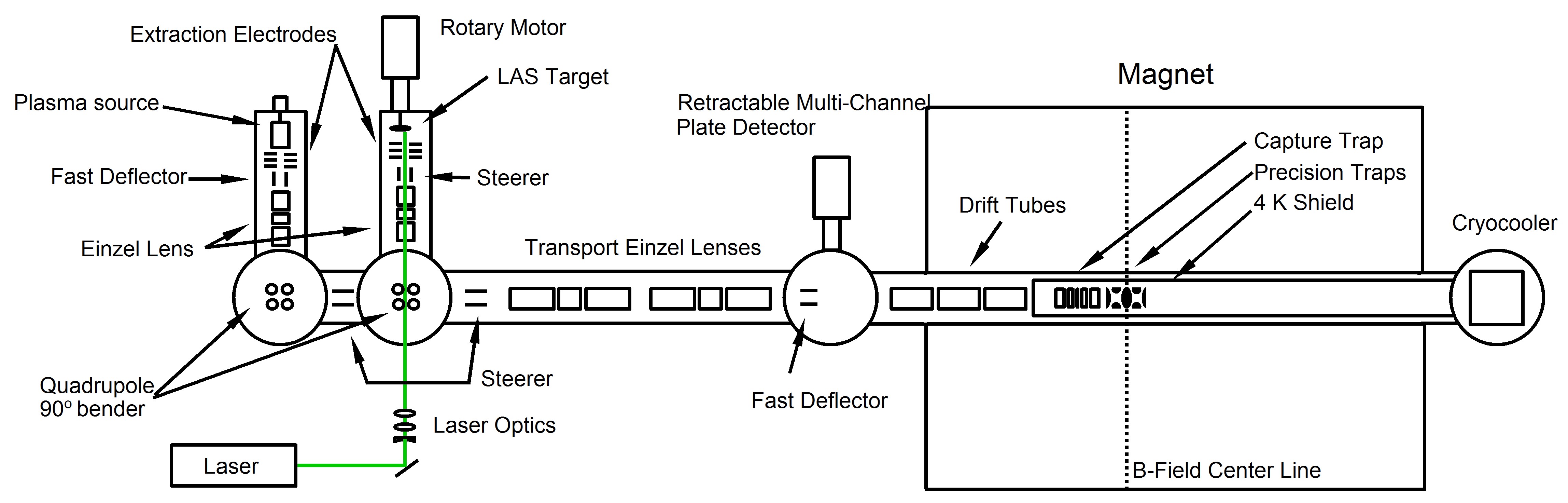}
\caption{\label{Layout}Schematic of CHIP-TRAP experimental layout.}
\end{figure}

The Central Michigan University high-precision Penning trap (CHIP-TRAP) is being designed with the goal of performing high-precision mass measurements on long-lived radioactive isotopes. It will combine the external ion production and transport of on-line TOF-ICR Penning trap facilities with the ultra-high precision, single-ion sensitivity of IC detection Penning traps. Ions will be produced with a plasma ion source and a laser ablation ion source (LAS). A schematic of the experimental layout is shown in Fig. \ref{Layout} The LAS, in particular, enables access to a wide range of isotopes and minimizes the amount of radioactive source material required.\footnote{Measurements performed using the destructive TOF-ICR and PI-ICR techniques have used as few as 10$^{15}$ atoms e.g. see Refs. \cite{Eliseev2015,Eibach2014}; since CHIP-TRAP will use a non-destructive measurement technique, the number of required atoms should be able to be reduced by 2 or 3 orders of magnitude.} Transport of the low-energy ions from the ion source will enable time-of-flight discrimination of non-isobaric contaminant ions \cite{Schury2006}, while Fourier Transform Ion Cyclotron Resonance (FT-ICR) techniques will be employed in a capture/filter trap to enable the identification and removal of isobaric contamination. A pair of precision traps housed in a 12 T superconducting magnet will be used for simultaneous cyclotron frequency comparisons to reduce the effect of magnetic field fluctuations. The traps will be operated at LHe temperatures to provide an ultra-high vacuum for long measurement times, and to minimize thermal noise in the resonant detection circuit.

\subsection{Laser ablation ion source}
Laser ablation ion sources have been installed and operated at a number of on-line Penning trap facilities to provide a convenient source of reference ions, such as carbon clusters, and to enhance off-line measurement capabilities, e.g. see Refs. \cite{Blaum2002,Chaudhuri2007,Elomaa2008,Smorra2009,Izzo2015}.

CHIP-TRAP employs a Continuum Surelite II, frequency doubled, pulsed Nd:YAG laser that can deliver up to 160 mJ per 5 ns pulse with a repetition rate of up to 20 Hz. The minimum beam diameter, occurring approximately 5 cm after the exit port of the laser, is 6 mm. For laser ablation, energy densities of ~10$^{8}$W/cm$^{2}$ are necessary. Hence, the laser is focused to a spot diameter of $<$1.0 mm using a three lens system \cite{Smorra2008}. Targets will be mounted on a rotatable circular target holder connected to a motorized vacuum feedthrough. This increases the lifetime of the target and enables multiple isotopes to be studied by ablating multiple targets attached to the holder.

\begin{figure}
\centering
\includegraphics[scale=0.4]{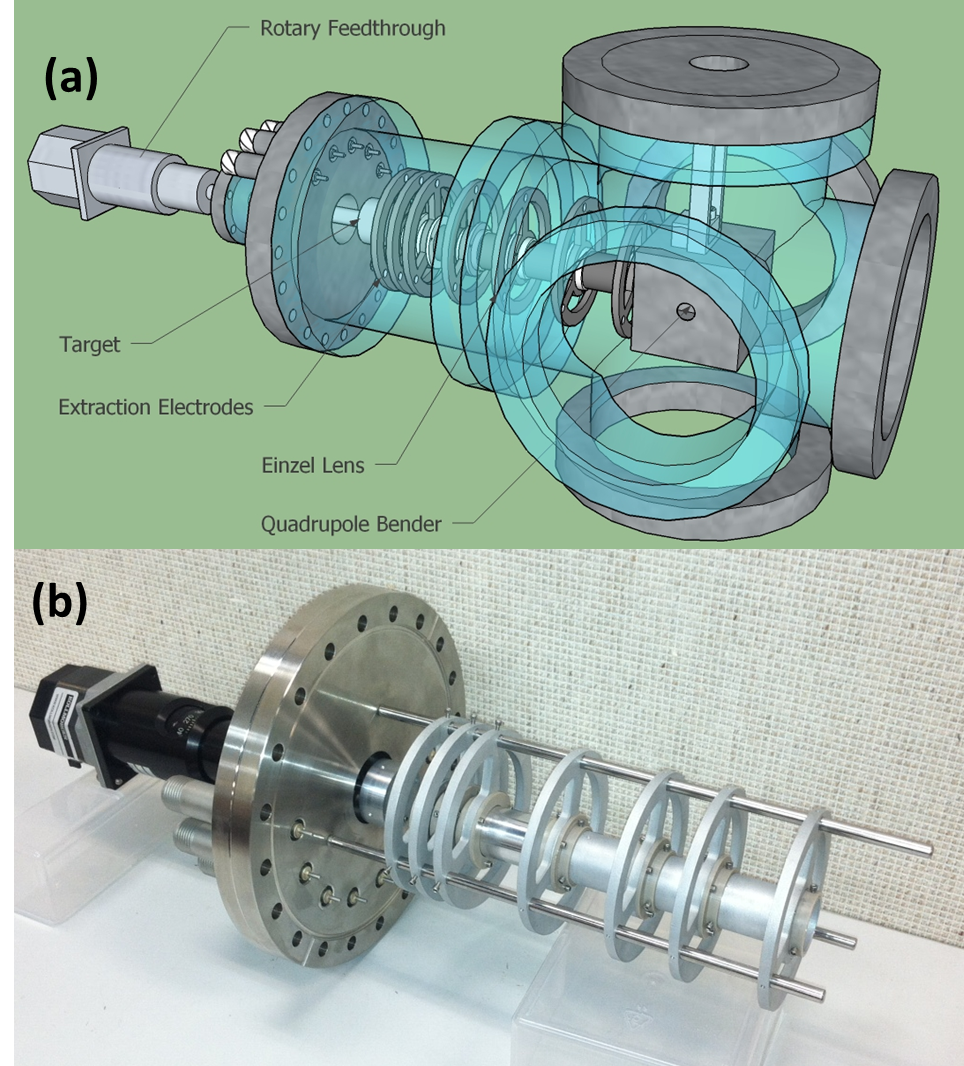}
\caption{\label{LAS_CAD_Pic}(a) CAD drawings of the LAS components and vacuum housing, (b) Photograph of fabricated LAS electrodes and rotary feedthrough mounted to an 8" conflat flange.}
\end{figure}

Ions that are ablated from the surface of the target are extracted, accelerated to 1 keV, and focused by a series of electrodes. Next they enter a 90$\degree$ bender and are directed into the main transport beamline. The 90$\degree$ bend separates neutral atoms produced by laser ablation from the ion beam. The design of the LAS extraction optics was based on Ref. \cite{Smorra2008} and was optimized by performing ion transport simulations using SIMION. The mechanical design was performed using the 3D CAD software SketchUp, see Fig. \ref{LAS_CAD_Pic}a, and the electrodes were fabricated in the CMU machine shop, see Fig. \ref{LAS_CAD_Pic}b.

\subsection{Ion transport and capture simulations}
\begin{figure}
\centering
\includegraphics[scale=0.6]{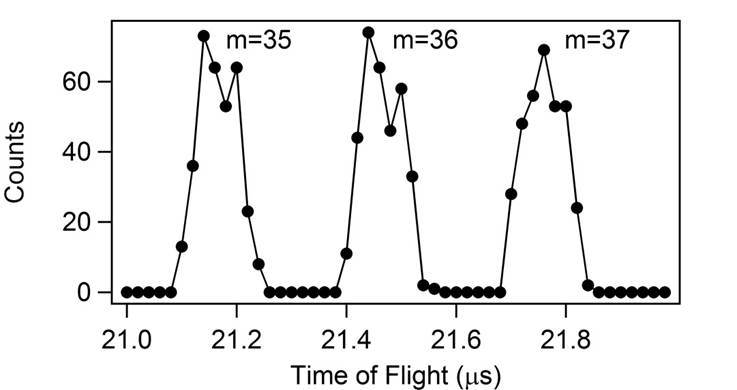}
\caption{\label{Cl_TOF}Results of simulations showing the time of flight of ions from the LAS to the location of the MCP. All ions leave the LAS with an energy of 1 keV and are hence temporally separated when they arrive at the MCP a distance of $\sim$1 m away.}
\end{figure}

After the quadrupole bender, ions are transported a distance of $\sim$1 m and focused onto a multi-channel plate (MCP) detector by a pair of einzel lenses. Simulations indicate that a 1 mm diameter circular beam can be produced on the MCP. The electrode in front of the MCP will be used as a fast electrostatic gate to deflect ions and open at a well-defined time to allow through ions of a given $m/q$ ratio. Fig. \ref{Cl_TOF} shows simulated results of the temporal separation of chlorine isotopes with an energy of 1 keV at the location of the MCP. 

\begin{figure}
\centering
\includegraphics[scale=0.6]{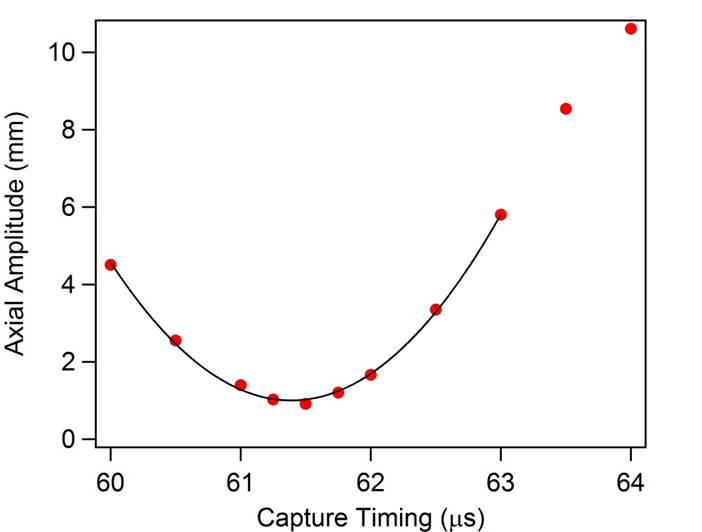}
\caption{\label{CaptureAmp}Results of simulations in which the time between ion creation and capture in the capture/filter trap was varied, and the resulting average axial amplitude of the ions in the trap was recorded.}
\end{figure}

Fast, but achievable switching times of $\sim$100 ns should enable isobaric separation. After the fast electrostatic gate, ions are focused into the fringe field of the magnet and pass through a series of drift tubes to reduce their energy so that they can be captured in the FT-ICR capture trap. The ions are captured by lowering the potential of the end-cap on the injection side of the capture trap and then raising it once the ions are close to the center of the trap. Fig. \ref{CaptureAmp} shows the results of simulations in which the capture time---the time from when the ions were created at the LAS to when the capture trap was closed---was varied, and the average axial amplitude was extracted. For the optimal capture timing, axial oscillation amplitudes of $\approx$1 mm were obtained.

The capture trap is a cylindrical geometry, orthogonally compensated closed end-cap Penning trap. The ring electrode is quarter-fold segmented. The end-caps have 4 mm diameter holes in the center, which, simulations show, enable 99\% capture efficiency. Once trapped, the FT-ICR technique will be applied to identify ions in the trap. Briefly, a broadband frequency sweep excitation is applied across two opposing segments of the ring electrode, followed by broadband detection of the image currents induced in the other two ring segments, which are connected to a low noise differential amplifier. The amplifier output is sampled and a fast Fourier transform (FFT) is performed on the digitized signal. Peaks, corresponding to the cyclotron frequencies of ions of different $m/q$ ratios, are observed in the frequency domain spectrum. By calibrating the magnetic field with ions of a well-known mass, the ion of interest can be identified. In subsequent capture cycles, contaminant ions can be removed from the trap by applying, for example, the SWIFT cleaning scheme \cite{Kwiatkowski2015,Guan1996}. The ion of interest can then be transported to either one of the two precision measurement traps. Simulations show that transport can be achieved by lowering the potential on the rear end-cap of the capture trap and the front end-cap of the first precision trap and then closing the precision trap when the ion is at the center (and similarly for the second precision trap).

\subsection{Penning trap design studies}
The geometry of the CHIP-TRAP precision measurement traps is based on that of the MIT-FSU Penning trap. Each trap has hyperbolic ring and end-cap electrodes and a pair of correction ring electrodes that can be used to null the lowest order $C_{4}$ term in the expansion of the electrostatic potential near the center of the trap \cite{Brown1986}:

\begin{equation}
\centering
   \Phi(\vec{r}) = \frac{V_{0}}{2}\displaystyle\sum_{n=2}^{even} C_{n}\left(\frac{r}{d}\right)^nP_n(\rm{cos}\theta)
\end{equation}

 The trap geometry was simulated using SIMION in order to characterize the electrostatic potential produced in the trap, and the effect of varying the potential applied to the correction ring, $V_{cr}$. As a check, the MIT-FSU trap was first simulated and studied with SIMION. For a given $V_{cr}$ setting, $\Phi(z)$ was extracted and Eqn. (1) was used to fit the potential up to $n = 8$ and the $C_{n}$ coefficients were extracted. $V_{cr}$ was then changed (with the ring and end-cap potentials remaining fixed) and the fitting procedure was repeated. A comparison of the simulated and measured parameters characterizing the MIT-FSU trap are given in Table I along with simulated results for CHIP-TRAP. $V_{cr0}$ is the potential that must be applied to the correction ring electrodes in order to null $C_{4}$\footnote{The MIT-FSU trap was designed so that nominally $V_{cr0}$ = 0. However, the effect of charge patches results in an offset from zero.}, $C_{6}$ is the next lowest order term, which cannot be tuned with the correction ring electrodes, $D_{4}$ is the gradient of $C_{4}$ vs $V_{cr}$. 

\begin{table}[h!]
\begin{center}
\begin{tabular}{ |c|c|c|c| } 
 \hline
  & Simulation & Experiment & CHIP-TRAP \\ 
 \hline\hline
 $V_{cr0}$ & -4(2) mV & 82(1) mV & -21(1) mV\\ 
 $C_{6}$ & 0.00072(62) & 0.00112(12) & -0.00038(7)\\ 
 $D_{4}$ & -0.0813(1) & -0.0821(42) & -0.0810(8)\\ 
 \hline
%\caption{\lable{Tab1} Table 1...
\end{tabular}
\caption{Comparison of simulated and experimental results characterizing the trapping potential of the MIT-FSU Penning trap and simulated results for CHIP-TRAP.}
\label{table:1}
\end{center}
\end{table}

The reduced cyclotron frequency of ions in the precision traps will be measured using the MIT pulse and phase (PNP) technique \cite{Cornell1989}. In this measurement technique only the axial motion of the ion is directly detected with a resonant LCR circuit. The reduced cyclotron frequency is measured by phase coherently coupling the radial and axial mode using a tilted quadrupole rf electric field. This enables a measurement of the axial frequency, $f_{z}$, and the reduced cyclotron frequency, $f_{+}$. The magnetron frequency, $f_{-}$, can be obtained separately, for example from an avoided crossing measurement \cite{Cornell1990}. These three frequencies can then be combined using the Brown-Gabrielse invariance theorem \cite{Gabrielse1982} $f_{c}^{2} = f_{+}^{2} + f_{z}^{2} + f_{-}^{2}$, to obtain the true cyclotron frequency, $f_{c}$.

CHIP-TRAP aims to perform simultaneous cyclotron frequency measurements on pairs of single ions confined in the two precision traps. This will greatly reduce one of the main sources of statistical uncertainty in cyclotron frequency ratio measurements: temporal magnetic field fluctuations. However, since the two ions are at slightly different locations in the magnetic field, a systematic shift in the cyclotron frequency ratio will be introduced. This can be eliminated by swapping the ions between the two traps and repeating the measurement. Using our SIMION model, we have simulated ion swapping between the two traps by lowering the potential of the shared end-cap for a well-defined period of time, and then raising the shared end-cap potential back to its original value. The time for the ions to switch between the traps is $\sim$10 $\mu$s. The residual axial amplitude after swapping the ions depends on the time taken to raise and lower the end-cap potential. Rise and fall times of $\leq$1 $\mu$s result in axial amplitudes of a few hundred $\mu$m.

\section{Summary and outlook}
CHIP-TRAP is a novel, double Penning trap currently being designed and constructed at Central Michigan University with the goal of performing high-precision mass measurements on long-lived and stable isotopes. A laser ablation ion source will be utilized to minimize the amount of source material required. Simultaneous cyclotron frequency comparisons on pairs of single atoms in the two traps will be performed in order to reduce the effect of magnetic field fluctuations to reach ultra-high precisions ($\sim$10$^{-11}$).

The LAS has been designed following ion transport simulations performed with SIMION and the electrode structure has been fabricated and assembled. A 532 nm pulsed Nd:YAG laser and necessary optics have been utilized to produce a beam with sufficient power density at the location of the target for laser ablation. Construction of the LAS is underway and tests of ion production will begin shortly. 
Simulations of ion transport from the LAS to the capture trap and precision measurement traps inside the 12 T magnetic field have been completed. These simulations indicate the ions can be transported and captured with high-efficiency using fast but relatively simple capture schemes. The design of the transport electrodes and beamline in CAD software is currently underway.
The Penning traps have been designed using SIMION to model the electrostatic potentials inside the trap. Electric field imperfections are expected to be of similar magnitude to those of other precision Penning traps, e.g. the MIT-FSU trap. Simulations of a simple swapping scheme in which the shared end-cap potential is quickly pulsed indicates that the ions can be switched between the traps without creating large axial oscillations. Detailing for the mechanical design of the traps is in progress with fabrication to follow.

\section*{Acknowledgements}
Funding was provided in part by Central Michigan University, Michigan State University and the Facility for Rare Isotope Beams, and the National Science Foundation under Contact No.1307233. We acknowledge technical assistance from Ray Clark and Mark Wilson.

\section*{References}

\bibliography{CHIPTRAPrefs}
%\begin{thebibliography}{00}
%\bibitem{Krempel2010} Krempel.

\end{document}